\documentclass{webofc}
\usepackage[varg]{txfonts}   
\usepackage{subcaption}

\begin{document}
\title{KEK Accelerator Test Facility Low-Level RF and Timing Systems}
%
%

\author{\firstname{Konstantin} \lastname{Popov}\inst{1, 2}\fnsep\thanks{\email{popovkon@post.kek.jp, constantinpopov79@gmail.com}} \and
        \firstname{Alexander} \lastname{Aryshev}\inst{1, 2}\and
        \firstname{Hiroshi} \lastname{Kaji}\inst{1, 2}\and
        \firstname{Toshiyuki} \lastname{Okugi}\inst{1, 2}}

\institute{High Energy Accelerator Research Organization (KEK), 305-0801 Tsukuba, Japan \and The Graduate University for Advanced Studies (SOKENDAI), 240-0193 Kanagawa, Japan}

\abstract{%
The KEK Accelerator Test Facility (ATF) is a dedicated testbed for nanobeam technologies in support of the International Linear Collider (ILC). Stable pulsed operation requires synchronization of the facility timing system with the Low-Level RF (LLRF) system. The timing system distributes trigger and gate signals to key subsystems, including the DAQ, klystrons, laser systems, pulsed kicker magnets, and interlocks. The LLRF system provides phase-coherent RF references and facility-wide clock distribution for synchronization. Achieving \textasciitilde100 fs-level synchronization depends critically on the phase-noise power spectral density (PN-PSD) of the distributed clock signals and on preserving this performance throughout the distribution network. We present facility-wide measurements of the KEK ATF LLRF clock PN-PSD and discuss the resulting synchronization floor imposed by the stability of the ATF Linac and Damping Ring signal generators.  
}
\maketitle
\section{Introduction}
\label{intro}
Precise synchronization between facility trigger and clock signals \cite{LLRF_10fs_RMS, Spring8_WRS_clock, Kaji_WRS_SuperKEKB} is a prerequisite for nominal operation of a pulsed accelerator. At KEK ATF, this synchronization is provided by an event generator/event receiver system locked to the accelerating-field sub-harmonics. It is complemented by a linear synchronizer \cite{Popov_ATF_LLRF, Popov_ATF_timing, Kaji_ATF_timing}. Beyond event timing, the phase stability of the low-noise Radio-Frequency (RF) reference \cite{Spring8_LLRF} and Local Oscillator (LO) signals \cite{Popov_IDROGEN_1, Popov_IDROGEN_2} delivered to laser, klystrons, and DAQ branches is a primary determinant of overall synchronization performance. It sets the sub-picosecond stability of electron bunch arrival time and accelerating-field phase. For this reason, facility-wide characterization of the clock-distribution architecture is essential. The phase-noise power spectral density (PN-PSD) and the corresponding rms time jitter and cumulative rms time jitter over relevant offset-frequency ranges provide direct, quantitative LLRF synchronization metrics.
The KEK ATF LLRF and timing systems must maintain precise synchronization among multiple subsystems. These include the RF-gun laser pulse \cite{Terunuma_RF_Gun, Popov_RF_Gun}, Linac and Damping Ring (DR) high-power RF \cite{Popov_ATF_LLRF}, Final Focus (FF) cavity beam position monitors (cBPMs) \cite{cBPM_1, cBPM_2}. In the context of the ongoing KEK ATF upgrade \cite{ATF_goals}, we evaluated the PN-PSD of the LLRF clock-distribution branches across the entire facility. 
These measurements were performed with the Agilent E5052B Signal Source Analyzer (SSA) \cite{SSA_method_1, SSA_method_2, KEK_LUCX_proceed} in PLL carrier-tracking/direct-homodyne mode to establish a baseline of the present system and to identify practical directions for further performance upgrades. The SSA Intermediate Frequency (IF) gain was 50 dB. The data was averaged 20 times with 5 correlations. The offset frequency measurement range is from 1 Hz to 10 MHz, but the 10 MHz reference signal case is exception. The upper bound is 5 MHz.
\section{KEK Accelerator Test Facility Low-Level RF clock distribution system}
The KEK ATF LLRF is based on two Agilent E8663B Signal Generators (SGs). The first one is dedicated to the RF-Gun laser system, Linac, while the second SG provides signals to the DR, Extraction line (EXT) and FF.  The DR SG is synchronized with the Linac SG by the low-noise 10 MHz reference signal. Therefore, the Linac SG is the ATF LLRF system grandmaster. The clock signals are distributed from the SG LLRF station location to the local stations via \textbf{P}hase \textbf{S}tabilized \textbf{O}ptical \textbf{F}iber (PSOF) without active delay compensation \cite{LLRF_10fs_RMS}. The Linac SG generates 1428 MHz continuous wave (CW) signal, which is injected into the frequency multiplier\&divider module to get the 2nd harmonics equal to 2856 MHz and the 4th, the 8th sub-harmonics equal to 357 MHz and 178.5 MHz, respectively. The DR SG provides 714 MHz clock for DR RF system. Moreover, this clock is connected to the frequency ramp electronics for energy ramp in the ATF DR, which is necessary for the dispersion correction procedure. 
\subsection{KEK Accelerator Test Facility Linac branches}
The ATF Linac signal generator (SG) acts as the grandmaster of the facility-wide LLRF system. It provides clock signals to the RF-gun laser system, the Linac klystrons local LLRF stations, diagnostics re-synchronization modules, and the L-band cBPM study station \cite{KNU_cBPM}. The klystron and RF-gun laser clock signals are derived using a frequency multiplier/divider module. The klystron clock at 2856 MHz is then amplified. It is split into 9 branches by a power divider, matching the number of klystrons.
At the next stage, each klystron clock branch is transferred to its local station independently. This transfer is performed via E/O and O/E conversion.
To quantify phase stability, the PN-PSD was integrated from 1 Hz to 10 MHz to obtain the integrated rms time jitter. Figure 1a and 1b show the Linac LLRF branch PN-PSDs with the SG Frequency Modulation (FM) disabled and enabled, respectively \cite{Popov_ATF_LLRF}. The FM bandwidth is 10 kHz. The Linac SG RF output integrated rms time jitter is 70 fs with FM disabled. With FM enabled, it increases to 1.64 ps. These rms jitter levels are preserved through the up-conversion to 2856 MHz, subsequent power amplification, and splitting (see Fig. 1a and 1b). The E/O and O/E conversion chain changes the rms time jitter \cite{Popov_ATF_LLRF}. 
\begin{figure}[h]
\centering
\includegraphics[width=1.0\textwidth,clip]{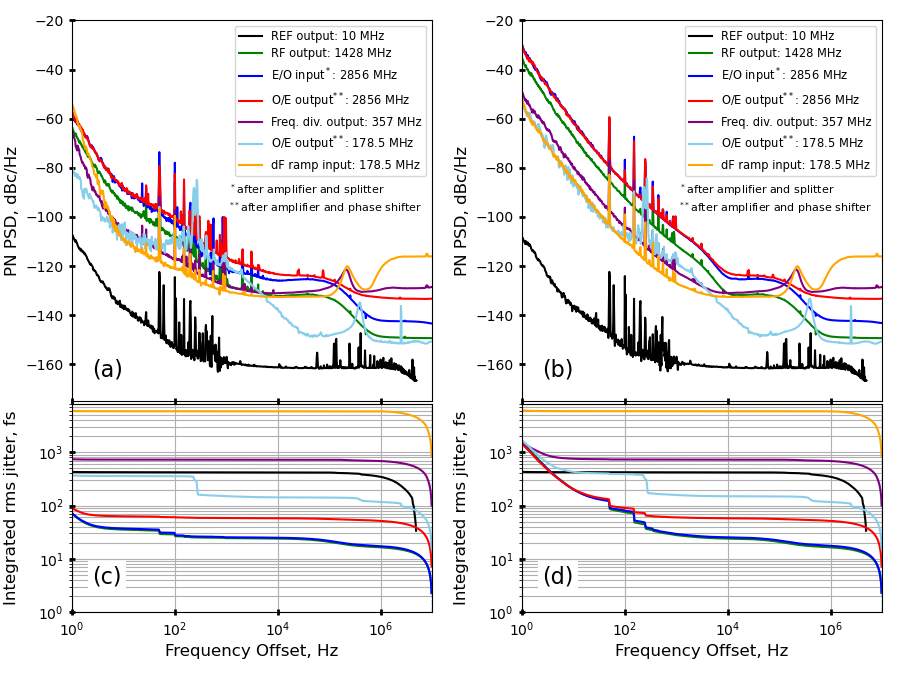}
\caption{The KEK ATF Linac LLRF branches: (a) is the PN-PSDs with disabled FM, (b) is the PN-PSDs with enabled FM, (c) is the cumulative integrated jitter (upper bounded at 10 MHz) with disabled FM, (d) is the cumulative integrated jitter (upper bounded at 10 MHz) with enabled FM}
\label{fig-1}       
\end{figure}
With FM disabled, it degrades the rms time jitter from 70 fs to 120 fs. With FM enabled, it improves the rms jitter from 1.64 ps to 1.45 ps. Table 1 summarizes the integrated rms time jitter for every klystron branch end point. These values differ from one klystron to another due to the different model of the E/O and O/E converters and LLRF feedback setup presence \cite{Popov_ATF_LLRF}. For readability, Table 1 reports the integrated rms time jitter in fs for the nominal mode (FM OFF) and in ps for the operational mode with frequency modulation (FM ON), reflecting the order-of-magnitude increase when FM is enabled.
\begin{table}[h]
  \centering
  \caption{The KEK ATF Linac klystrons LLRF station end-point integrated rms time jitter}
  \label{tab:mytable}
  \begin{tabular}{cccccccccc}
    \hline
    \rule[-1.9ex]{0pt}{4.8ex}\parbox[l]{0.7cm}{\centering Klystron\\number} & \parbox[c]{0.7cm}{\centering 0} & \parbox[c]{0.7cm}{\centering 1} & \parbox[c]{0.7cm}{\centering 2} & \parbox[c]{0.7cm}{\centering 3} & \parbox[c]{0.7cm}{\centering 4} & \parbox[c]{0.7cm}{\centering 5} & \parbox[c]{0.7cm}{\centering 6} & \parbox[c]{0.7cm}{\centering 7} & \parbox[c]{0.7cm}{\centering 8} \\
    \hline
    \parbox[c]{2.2cm}{\centering \rule{0pt}{2.2ex}RMS jitter\\(FM OFF)$^{1}$, fs} & 120 & 120 & 160 & 150 & 120 & 120 & 150 & 190 & 90 \\
    \parbox[c]{2.2cm}{\centering \rule{0pt}{2.2ex}RMS jitter\\(FM ON)$^{2}$, ps}  & 1.45 & 1.58 & 1.68 & 1.58 & 1.60 & 1.52 & 1.72 & 1.58 & 1.48 \\
    \hline
  \end{tabular}
\vspace{-2pt}
\begin{minipage}{\linewidth}
\footnotesize\raggedright
\textsuperscript{1} FM OFF: Frequency Modulation is disabled.\\
\textsuperscript{2} FM ON: Frequency Modulation is enabled.
\end{minipage}
\end{table}

After the same E/O and O/E conversion, the RF-gun laser clock rms time jitter is 370 fs with FM disabled. With FM enabled, it is 1.71 ps.
The branch serving the frequency-ramp electronics shows a much larger phase-noise degradation in both modes. The integrated rms time jitter is increased to 6.02 ps with FM disabled. It is 6.15 ps with FM enabled. A similar degradation is observed for the 357 MHz clock signal. With FM disabled, the integrated rms time jitter is 750 fs. In the other mode, it is increased to 1.6 ps. The only exception is SG 10 MHz reference output port. Its rms jitter is always equal to 430 fs, which is integrated to 5 MHz. Figure 1c and 1d show the cumulative integrated rms time jitter bounded at 10 MHz offset frequency. The jitter for FM disabled mode is limited by the SG electronics itself and frequency conversion modules. With FM enabled, the modulation bandwidth contribution becomes dominant at PN PSD (see Fig 1d).
\subsection{KEK Accelerator Test Facility Damping Ring branches}
The ATF DR clock is passed through the frequency ramp electronics set, which includes frequency divider, LO generator, down-converter, phase monitor and feedback controller. Also, the SG 10 kHz frequency modulation is enabled in order to compensate the phase unlocking during the frequency ramp operation \cite{Popov_ATF_LLRF}. If the frequency ramp electronics is bypassed and the FM is disabled, the DR SG output port integrated rms time jitter is equal to 90 fs for 714 MHz (see Fig. 2a and 2c). When the frequency ramp electronics with associated PLL operates together with the FM, the integrated rms time jitter is equal to 1.99 ps (see Fig. 2b and 2d). Then, the signal is transferred to the Damping Ring local LLRF station via heliax cable. The rms time jitter of 714 MHz is equal to 90 fs after the transfer, amplification and splitting, when FM is disabled. However, it degrades from 1.99 ps to 2.02 ps for enabled FM. The ATF DR LLRF analog phase\&amplitude feedback further degrade the rms time jitter from 90 fs to 1.09 ps, due to the feedback characteristics. It is the input to the DR klystron drive amplifier. 
\begin{figure}[h]
\centering
\includegraphics[width=1.0\textwidth,clip]{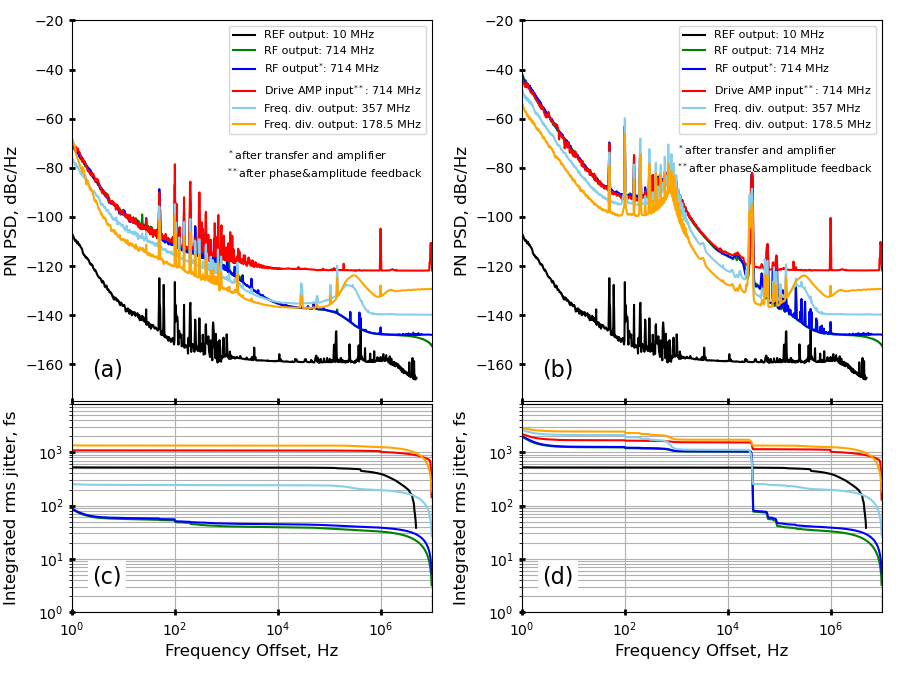}
\caption{The KEK DR LLRF branches: (a) is the PN-PSDs with disabled FM, (b) is the PN-PSDs with enabled FM, (c) is the cumulative integrated jitter (upper bounded at 10 MHz) with disabled FM, (d) is the cumulative integrated jitter (upper bounded at 10 MHz) with enabled FM}
\label{fig-1}       
\end{figure}
The cumulative integrated rms jitter (see Figure 2c) shows that DR SG output port rms time jitter is dominated by its own PLL equipped with OCXO. However, Figure 2d depicts the sudden drop after 10 kHz frequency offset, which clearly states that external PLL associated with the frequency ramp is the dominant contributor to the integrated rms time jitter. Also, the DR SG output signal is down-converted from 714 MHz to 357 MHz and 178.5 MHz by single module. The DR SG output 2nd sub-harmonics rms time jitters are equal to 250 fs and 2.64 ps for disabled and enabled FM. The noises associated with 178.5 MHz output port are equal to 1.36 ps and 2.92 ps (see Fig.2a and Fig.2b). The DR SG 10 MHz reference output demonstrate the noise equal to 520 fs, which is integrated to 5 MHz. The DR SG 10 MHz reference rms time jitter is higher than the Linac SGs one because of the daisy chain processing inside the SG electronics.
\subsection{KEK Accelerator Test Facility Final Focus branches}
The DR RF clock signal equal to 714 MHz processed by the LLRF phase\&amplitude feedback is sent to the ATF FF local LLRF station via optical fiber. Then, the signal is amplified to be injected into the frequency up-conversion module associated with the ATF FF cBPM Low-Level RF front-end electronics \cite{Popov_ATF_LLRF}. The integrated rms time jitter of the input to the up-conversion module signal degrades from 90 fs to  5.67 ps with disabled FM due to E/O and O/E conversion chain (see Fig. 3a and 3b). With FM enabled, the noise is increased from 2.02 ps to 5.67 ps. It up-converts the input signal from 714 MHz to 6.453 GHz. The C-band LO 6.453 GHz clock is utilized to down-convert the cBPM signal \cite{cBPM_1, cBPM_2}. 
\begin{figure}[h]
\centering
\includegraphics[width=1.0\textwidth,clip]{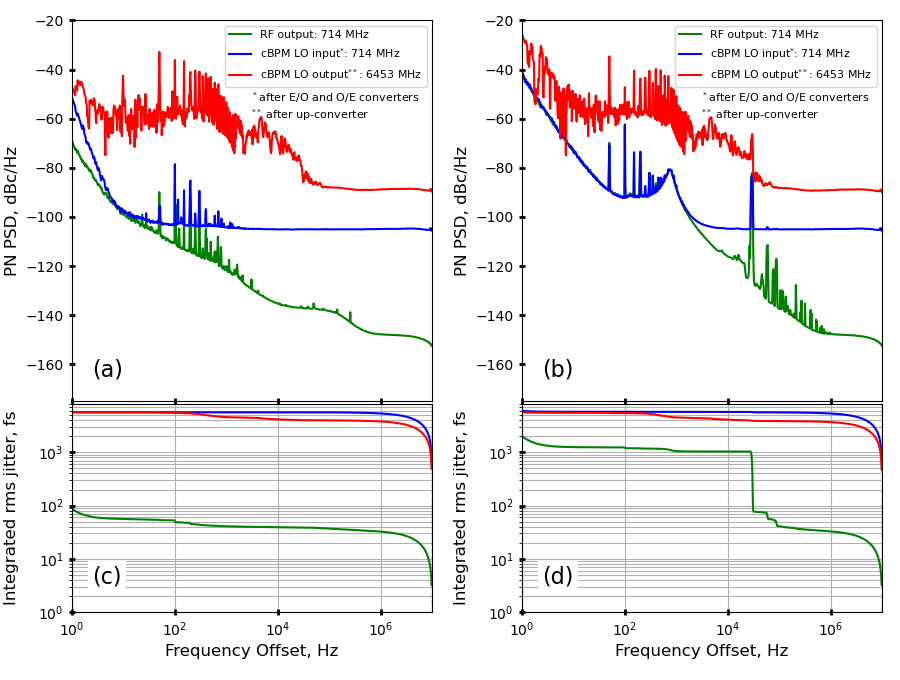}
\caption{The KEK ATF FF LLRF branches: (a) is the PN-PSDs with disabled FM, (b) is the PN-PSDs with enabled FM, (c) is the cumulative integrated jitter (upper bounded at 10 MHz) with disabled FM, (d) is the cumulative integrated jitter (upper bounded at 10 MHz) with enabled FM}
\label{fig-1}       
\end{figure}
The LO signal integrated rms time jitters are equal to 5.67 ps and 6.02 ps for FM disabled and enabled, correspondingly. The 6.453 GHz LO inherits the input jitter through the up-conversion chain. The cumulative integrated rms noises show the same pattern as 714 MHz signal (see Fig. 3c and 3d).
\section{Conclusion}
In this work, we evaluated the facility-wide phase-noise performance of the KEK ATF reference-clock distribution by measuring PN-PSD at key points of the LLRF network and converting the spectra to integrated rms time jitter over 1 Hz–10 MHz. For the Linac klystron reference at 2.856 GHz, the measured integrated jitter at the local stations is on the order of 100 fs, demonstrating that the RF distribution via E/O and O/E conversion preserves low-jitter performance when operated under nominal settings.

A significant degradation is observed when the DR reference is operated with frequency modulation and the ramp-generation loop is engaged. The PN-PSD level increases by approximately a few tens of dB over a broad offset-frequency range, resulting in integrated jitter at the level of a few picoseconds. This behavior is consistent with the DR synchronization architecture, in which the ramp-generation feedback loop and associated electronics dominate the phase-noise contribution and the resulting fluctuations propagate to downstream subsystems.

These results identify the DR ramp-generation path as the primary limitation for further facility-wide jitter reduction. Future upgrades should therefore focus on the ramp-generation feedback architecture and its implementation aspects, such as controller/noise shaping, loop bandwidth and actuator chain, and location of critical electronics, with the goal of extending sub-100 fs-class stability to all ATF sections under operational conditions.

\section{Acknowledgement}
This work was supported by the MEXT program “Development of key element technologies to improve the performance of future accelerators,” Japan Grant Number JPMXP1423812204. Authors thank Tetsuya Kobayashi for his help with Agilent SSA E5052B Signal Source Analyzer.
%
%
%

\end{document}